# The role of negative emissions in meeting China's 2060 carbon neutrality goal

Jay Fuhrman[1,2], Andres F. Clarens[1], Haewon McJeon[2], Pralit Patel[2], Yang Ou,[2] Scott C. Doney[3], William M. Shobe[4], Shreekar Pradhan[1]*

[1] Department of Engineering Systems and Environment, University of Virginia, Charlottesville, Virginia, USA

[2] Joint Global Change Research Institute, University of Maryland and Pacific Northwest National Laboratory, College Park, Maryland, USA

[3] Department of Environmental Sciences, University of Virginia, Charlottesville, Virginia, USA

[4] Batten School of Leadership and Public Policy, University of Virginia, Charlottesville, Virginia, USA

*Corresponding author, Department of Engineering Systems and Environment, University of Virginia, Charlottesville, Virginia, 22904, USA, email: shreekar@virginia.edu

April 15th, 2021

## Abstract

China's pledge to reach carbon neutrality before 2060 is an ambitious goal and could provide the world with much-needed leadership on how to limit warming to $+1.5^{O}C$ warming above pre-industrial levels by the end of the century. But the pathways that would achieve net zero by 2060 are still unclear, including the role of negative emissions technologies. We use the Global Change Analysis Model to simulate how negative emissions technologies, in general, and direct air capture (DAC) in particular, could contribute to China's meeting this target. Our results show that negative emissions could play a large role, offsetting on the order of 3 $GtCO_2$ per year from difficult-to-mitigate sectors such as freight transportation and heavy industry. This includes up to a 1.6 $GtCO_2$ per year contribution from DAC, constituting up to 60% of total projected negative emissions in China. But DAC, like bioenergy with carbon capture and storage and afforestation, has not yet been demonstrated at anywhere approaching the scales required to meaningfully contribute to climate mitigation. Deploying NETs at these scales will have widespread impacts on financial systems and natural resources such as water, land, and energy in China.

## 1. Introduction

On September 22, 2020 China announced that it would pursue a plan to achieve "carbon neutrality" in its economy by 2060.[1] China has previously committed to peaking its $CO_2$ emissions before 2030,[2] and its new carbon neutrality commitment greatly strengthens its nationally determined contributions (NDCs) under the Paris Climate Agreement. There are a number of studies that have explored China's decarbonization pathways,[3–8] but not much attention has been given to the role that negative emissions technologies (NETs) could play, especially the availability of direct air capture (DAC). In our recent paper, we assessed the negative emissions requirement for meeting a +1.5⁰C target globally.[9] Our results indicated that DAC could play a large role with much of that activity concentrated in the US and China because of their substantial capacity to carry out geologic carbon storage. [9]

The number of countries, regional governments, and corporations that have been making carbon neutrality commitments has been accelerating.[10] While China is distinct from many of the other countries and institutions making decarbonization pledges because of its size, its announcement was similar to other national commitments such as the United Kingdom's and Japan's that have provided few details about implementation and enforcement.[11–13] China today produces approximately one-third of global greenhouse gas emissions, but up to 30 percent of these emissions result from production of goods that are exported to other regions of the world.[14–18] In addition, energy consumption in China is currently highly carbon intensive with most of its primary energy supply coming from coal combustion.[19,20] Wang et. al., (2019a) examined pathways to decarbonize China's power sector, including the use of BECCS, and the early retirement of coal-fired generation units.[21] But China has recently invested in 38 GW of new coal capacity in 2020 and has indicated intention to build hundreds of new coal-fired power plants in its most recent 5-year plan.[22–24] This is at odds with its stated efforts to decarbonize, as well as those to improve air quality. As reported by Wang et. al., (2019b), China's late-stage industrial sector could rapidly reduce $CO_2$ emissions over the coming decades under global climate policy with a combination of energy efficiency improvements, fuel switching, and CCS. However, owing to its size relative to other large economies such as Western Europe, and the prohibitive expense of fully decarbonizing some industrial processes (e.g., iron and steel), China's industrial sector could have over 1 Gt-$CO_2$ of residual emissions by 2050.[25] Transportation in China and elsewhere is expected to remain recalcitrant to decarbonization relative to other sectors, owing to continued dominance of petroleum-derived liquid fuels, especially for freight.[26–28] Like other large economies, China's $CO_2$ emissions declined temporarily due to the COVID-19 pandemic. However, emissions in China and elsewhere are rebounding and are likely resume their historic growth trajectory without investment in lower carbon technologies and deliberate policy action to reduce them.[29,30] In order to achieve its carbon-neutrality target, China will need to rapidly decarbonize its power, transportation, and industrial sectors in the near term and will likely have to seek opportunities for negative emissions in the long term.[31,32]

To achieve carbon neutrality, a country needs to balance emissions and sinks. For any large and complex economy, there will be sources of emissions that will be recalcitrant to decarbonization, such as aviation, freight transport, and high temperature heat applications in industry. For this reason, there is growing interest in approaches for actively removing emissions from the atmosphere.[33] So-called negative emissions technologies (NETs) are a suite of engineered or natural approaches such as DAC, bioenergy with carbon capture and storage (BECCS), and afforestation that remove carbon from the atmosphere and could play an important role in offsetting recalcitrant emissions, and/or reaching net-negative emissions globally or regionally. China has attempted afforestation projects to combat desertification and soil erosion in the past, with mixed success in initial phases.[34–36] Adapting forest protection and afforestation approaches based on lessons learned could reduce negative impacts on biodiversity and water availability, and some studies indicate that the expanding forest cover in China may be generating large carbon sinks from the atmosphere.[37] Preventing further deforestation and restoring previously deforested land to its natural state has potential environmental and human health co-benefits, in addition to storing carbon from the atmosphere.[38] However, China's experience with afforestation highlights the potential challenges of further expanding of this approach for large-scale climate mitigation due to measurement uncertainties inherent to natural and managed forest systems, competition with agricultural land demands, opposing goals of carbon storage and timber harvest, large land and water footprints, and the potential lack of permanence of forest carbon stocks in a warming world.[39,40] As China's government seeks to increase the standard of living for its citizens, large-scale bioenergy crop cultivation could also compete with food production, as well as environmental conservation objectives.[41–43] DAC is an emerging technology with a far lower land footprint than BECCS or afforestation, but large energy demand due to the thermodynamic unfavourability of separating atmospheric $CO_2$ at ~415 PPM.[44,45] With recent, more optimistic cost estimates for DAC, several large European and U.S. based companies have made investments in commercial DAC technologies, and still others have committed to using negative emissions including DAC to draw down their historical emissions from the atmosphere.[46–50] Given China's currently large emissions, its capacity for geologic storage,[51] and the large share of its carbon-intensive exports in the global economy, DAC could potentially play a large role in deep decarbonization there. Yet there has not yet been modeling performed to understand the role and tradeoffs of DAC and other negative emissions in China in achieving its recently-announced climate ambitions.

At a global scale, integrated assessment models (IAMs) have been used to explore deep decarbonization pathways.[52] IAMs incorporate economic, geophysical, demographic, and climate modules to study future policy scenarios. The International Panel on Climate Change (IPCC) uses a suite of IAMs to explore different scenarios and inform international commitments, including those laid out in the 2015 Paris Agreement.[53] Over the past several years, IAMs have been used to explore what it would take to limit future anthropogenic climate change to $+1.5^{\circ}$C warming relative to pre-industrial levels. All the IAMS used by the IPCC show that, in order to meet these aggressive decarbonization scenarios, NETs will be needed to help offset recalcitrant emissions. BECCS and afforestation are the most widely modeled technologies with median projected global deployments of respectively, 4 and 2 Gt $CO_2$ per year, projected by 2050 to limit warming to below 2 C in 2100.[54] Several recent modeling studies have also assessed engineered NETs such as DAC, with even higher projected deployments to actively draw down atmospheric $CO_2$ levels.[55–58] Such large deployments of these NETs will entail enormous transitions for energy and land, as well as water use patterns, and it is critical to understand how these might unfold at the country-level.[9]

The goal of this paper is to estimate the potential role of DAC and other forms of negative emissions to help China meet its carbon neutrality goal, as well as their interactions with mitigation efforts. In particular, we examine how the availability of different NETs might affect the required decarbonization of different sectors and the extent to which each type of NET could be deployed. We estimate the costs and tradeoffs for land, water, and energy systems of negative emissions deployment in China. The scale at which they could be needed in order to meet this target is quite large, so it is crucial to understand the tradeoffs these technologies would represent. These results could also provide baseline cost estimates to inform where to target investments in innovation. To perform this analysis, we used the Global Change Analysis Model (GCAM)[59], a technology-rich integrated assessment model with embedded simplified versions of global climate and carbon dynamics. We modeled three main scenarios featuring estimates for DAC cost and energy intensity to assess how China might achieve its carbon neutrality target in 2060. In addition, we conducted a sensitivity analysis to understand how key model assumptions could influence the projected role of DAC in China. These results provide insight about the technological transitions, as well as financial, environmental, and human health impacts that could result from NETs deployments in China.

## 2. Methods

We used the latest release of GCAM 5.3[59], enhanced with the capability to model DAC, to simulate different paths by which China individually, as well as the rest of the world, collectively might reach net-zero emissions by 2060. We followed the near-term to net-zero ("NT2NZ") approach described in Kaufman et al.[60] We assumed a linearly declining net $CO_2$ emissions trajectory from 2021 until reaching zero net $CO_2$ emissions in 2060 for both China separately and the rest of the world together. This modeling approach for net-zero $CO_2$ emissions pathways accommodates uncertainties and measurement difficulties while also helping guide near-term policy design. Specifically, while the prospective future availability of DAC and other NETs would tend to delay mitigation in modeling scenarios where the $CO_2$ price rises at an assumed interest rate, this could prove to be a risky bet for real-world decision makers if NETs and other uncertain technologies prove unable to scale up as rapidly as expected, and/or as impacts from climate change itself worsen.[60–62] With this $CO_2$ constraint, we evaluated three parametrizations for the cost, efficiency and availability of DAC (i.e., high-cost DAC, low-cost DAC, and no DAC available). Population and GDP input assumptions follow the "middle of the road" scenario widely used in integrated modeling studies and can be found in the Supplementary Information.[63] Prices on land-use change and correspondingly, subsidies for afforestation, are specified as an increasing proportion of the fossil and industrial carbon emissions price, reaching 100% of the fossil carbon price by 2100. This is intended to represent gradually improving institutions for pricing land-use change emissions, given that agriculture and land-use decisions now largely occur outside of regulatory frameworks in most countries.[9,64]

Table 1. Description of the $CO_2$ emissions constraint trajectories run in GCAM to understand the role DAC and other NETs in meeting these constraints

We modeled DAC as a process that uses an aqueous reaction between atmospheric $CO_2$ and a hydroxide solution that has evaporative water losses at the air contactor.[65–67] The DAC technology requires energy input in the form of high-temperature process heat and electricity, and financial inputs for capital expenditure and non-energy operations and maintenance, given in Table 2. For low-cost DAC, financial and energy inputs are assumed to decline linearly between their 2020 and 2050 parametrizations, then remain constant after 2050. In the high cost DAC cases, the technology is assumed to remain costly and energy-intensive over time. We assume that the process heat is high-temperature heat from natural gas combustion and not lower temperature waste-heat or renewables. While there are other DAC archetypes that can use renewable electricity and/or waste heat input and do not consume water,[68] we focused our analysis on this high temperature process because it appears to be the most inexpensive and commercially mature at present.[65,69] Geological carbon storage costs are treated endogenously by GCAM. Note that DAC is assumed to behave as a quasi-backstop technology, with no external constraints on its deployment outside the availability of energy, carbon storage, and its cost relative to other mitigation and negative emissions technologies in meeting a binding cap on $CO_2$ emissions. No constraints were imposed on the scaling rate of DAC or any other technology. We assumed a median lifetime for DAC plants of 40 years. No other technological, institutional or legal limitations are modeled with respect to DAC, which we assume can be deployed rapidly at scale in the model under appropriate conditions. Because of these assumptions, the simulated rate of DAC deployment at the costs specified may be considered as an upper bound.

Table 2. Input parameters for DAC technology[56,65]

## 3. Results

Figure 1 presents the results for (a) global average temperature anomaly, (b) atmospheric $CO_2$ concentration, and (c) $CO_2$ emissions over time for China as well as globally. The $CO_2$ emission trajectories after 2020 follow directly from the constraints imposed. If China and the rest of the world together reduce their emissions to net-zero by 2060, this results in approximately $+1.8$ºC of warming in 2100. Scenarios in which DAC is deployed show slightly higher warming in 2100 despite meeting the same $CO_2$ emissions cap, owing to fugitive methane emissions from the production of natural gas which DAC takes as an input, as well as higher residual non-$CO_2$ emissions from difficult-to-mitigate sectors. In all scenarios, criteria air pollutant emissions in China (e.g., black carbon, VOC, NOx, and $SO_2$) are projected to decline drastically from their current levels due to the phase-down of coal as an energy source, regardless of the availability of DAC. Relative to their 2010 levels, black and organic carbon particulate emissions each decline by over 80%, NOx emissions decline by 73%, and $SO_2$ emissions decline by 95%, highlighting important environmental and public health co-benefits of $CO_2$ emissions reduction policy. Detailed projections for non-$CO_2$ emissions for our three main scenarios are provided in the Supplementary Information.

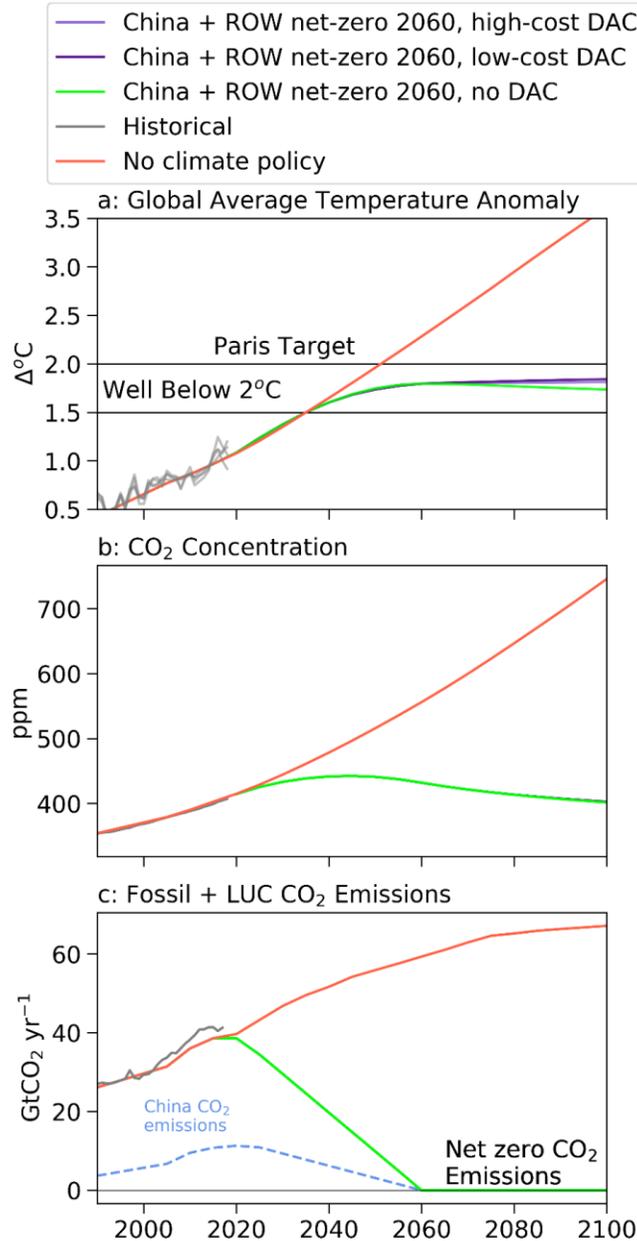

Figure 1: Model results show different trajectories for global temperature anomaly (a), even though $CO_2$ concentration (b), and net $CO_2$ emissions (c) over time are the same for the 3 scenarios simulated here. Globally net-zero emissions could limit end-of-century warming to below +2°C, but more ambitious near-term mitigation and/or future net-negative emissions is required to meet a below +1.5°C goal. **Historical data for emissions,**[70] **$CO_2$ concentrations,**[71] **and temperature anomaly**[72] **are indicated by grey lines.** Four data sources were used to report recorded historical temperature anomaly; the darker grey line indicates overlap or especially high agreement between different observations, which were plotted with slightly increased transparency for clarity. Projected warming is approximately 0.1 °C higher in the low-cost DAC scenario owing to increased fugitive

emissions from the natural gas supply chain to power DAC itself, as well as higher levels of non-$CO_2$ residual emissions from other economic sectors.

Figure 2 illustrates the extensive transformation China will need to make to its economy in order to achieve its 2060 net-zero target. The panel on the left shows the dominant current emission sectors: transportation, industrial, and electric power. On the right, the panel shows the results of three permutations of the China + rest of world net-zero 2060 scenario; one in which no DAC is available, one in which DAC is available but continues to be expensive, and one in which DAC is available and gradually improves in cost and energy efficiency by 2050 (see Table 2). In all three cases, China will rely on significant deployment of negative emissions to achieve its net-zero target.

Our results show that getting China and the rest of the world to net-zero by 2060 without DAC, would require China to deploy at least 1 Gt-$CO_2$ per year of negative emissions from BECCS and afforestation, which is in line with a recent study by Yu et al.[31] Achieving the net-zero $CO_2$ target without DAC available, that is, relying only on BECCS and afforestation for negative emissions, would result in a marginal cost of over $800 per t$CO_2$ in 2060 ($2015 price). With DAC available, the world along with China could achieve net-zero $CO_2$ emissions by 2060 at much lower marginal costs, in the range of US$ 200 - 400 per t$CO_2$. This carbon price corresponds to the sum of the capture + carbon storage costs for DAC, both of which become gradually more costly over time as geologic storage reservoirs and energy resources are consumed. Full carbon emissions price paths are reported in the Supplementary Information.

Under the low-cost DAC deployment, China would require about 3 Gt$CO_2$ per year of negative emissions to reach net-zero by 2060, much of it coming from DAC, with the remainder coming from BECCS and afforestation. If DAC does not improve in cost and energy-efficiency, it would make up a smaller percentage of the negative emissions portfolio, which would still total 2 Gt$CO_2$ per year. Residual, positive $CO_2$ emissions from difficult-to-mitigate sectors require multiple Gt-$CO_2$ of negative emissions to offset them in all scenarios. In the no DAC scenario, the industrial sector becomes a net sink of $CO_2$ largely through hydrogen production from bio-feedstocks and CCS. Low-cost DAC availability allows up to 1 Gt-$CO_2$ emissions from industrial energy use in China to be offset at lower cost. Transportation is projected to be recalcitrant to decarbonization across all scenarios. While passenger transportation can decarbonize through electrification and hydrogen fuel cells, freight trucks using both liquid and natural gas fuels under the carbon policy result in large residual emissions. An electrified and/or hydrogen fuel cell freight truck fleet in China and elsewhere could substantially reduce the need to deploy NETs to offset their emissions. Transportation emissions are slightly higher in the DAC scenarios due to higher service demand because the lower carbon emissions prices lead to correspondingly lower fuel prices. There is also less need to switch from vehicles using petroleum-derived fuels to electric, natural gas, and hydrogen fuel cell vehicles if DAC is available to offset these distributed emissions. Detailed breakdowns of industrial and transportation $CO_2$ emissions in China, as well as transportation demand are made available in the Supplementary Information.

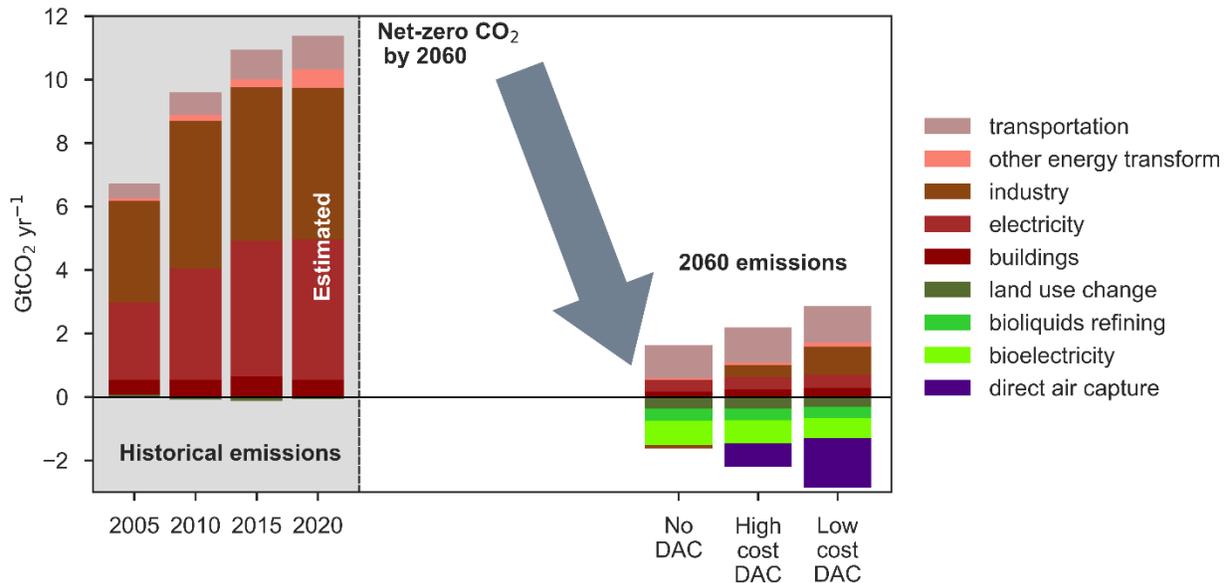

Figure 2: Pathways for China to reach net-zero $CO_2$ emissions by 2060 all involve deep emissions reductions and $CO_2$ removal. The availability of DAC technology in China enables less use of BECCS and also offsets emissions from difficult-to-mitigate sectors such as transportation and industry, allowing higher emissions from these sectors relative to the no DAC case. Results shown here are for the China + rest of the world (ROW) net-zero 2060 scenario, which results in approximately 1.8° C of warming from preindustrial in 2100.

Figure 3 shows the projected primary energy transitions for China to reach net-zero $CO_2$ emissions by 2060. At present, China relies heavily on coal and oil for its primary energy. To achieve net zero $CO_2$ emissions will require a substantial rollout of renewable energy over the coming 40 years in addition to very large deployment of fossil carbon capture and storage. For the three negative emissions cases described in Figure 3 for the China + rest of world net-zero by 2060 scenario (no DAC available, high cost, and low-cost DAC) we see important differences in energy consumption patterns. Notably, the deployment of DAC consumes natural gas for process heat on the same order of magnitude as all of China's present day gas consumption. The primary energy consumption of coal declines from 67% share in 2020 to 21% in 2060 without DAC. This includes 2% conventional coal and 19% coal-based carbon capture and storage. With low-cost DAC deployment, the share of coal would decline to 22% in 2060 with 4% conventional coal and 18% coal-based carbon capture and storage.

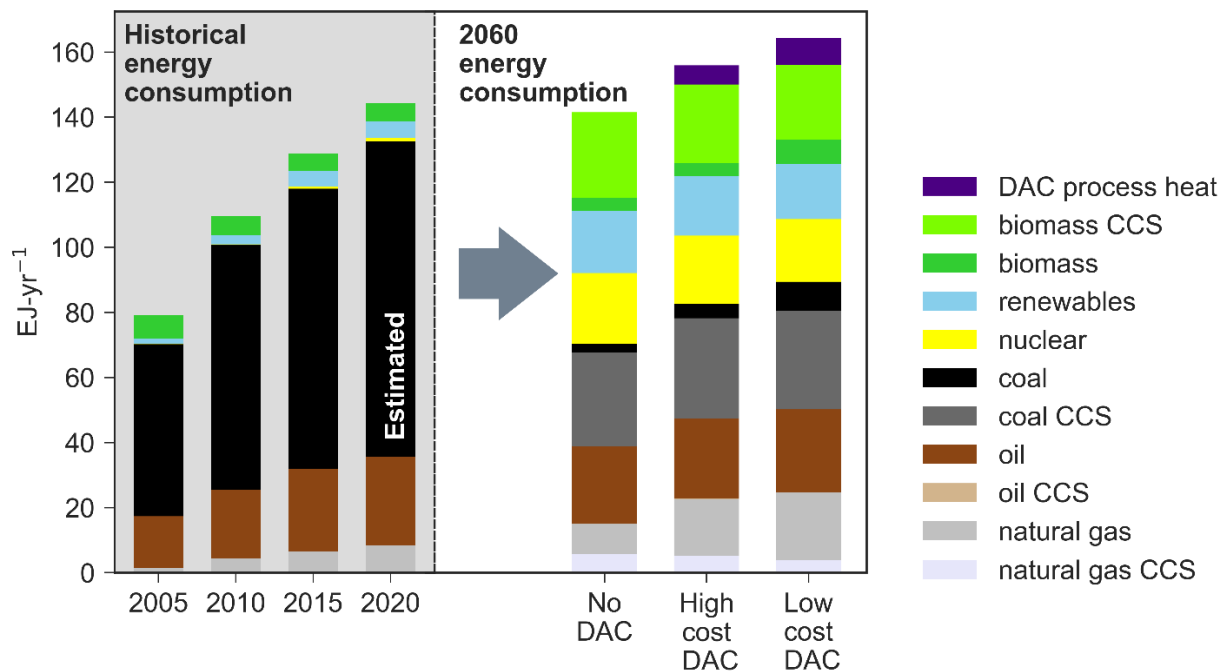

Figure 3: Historical and projected primary energy by fuel for China showing values for recent historical periods and for net-zero $CO_2$ emissions in 2060 scenarios. Process heat for DAC (i.e., the primary energy from natural gas CCS devoted solely to $CO_2$ removal) is reported separately in indigo (dark purple at top of bars) and subtracted from natural gas CCS to avoid double counting.

Figure 4 shows potential water consumption patterns for China in meeting its net-zero $CO_2$ emissions by 2060 goal. Irrigation for agriculture comprises most of the historical and projected water consumption, and evapotranspiration from rainfed agriculture is not included in this figure. Agricultural water consumption is projected to decline slightly by 2060 owing to technological improvements and exogenous assumptions of population peaking and then declining by mid-century. In all cases, irrigation for bioenergy crop cultivation expands from near-zero level in present day, to substantial fractions of overall water use in 2060. If no DAC is available, bioenergy crop irrigation grows to 7 km$^3$ per year in 2060, which is nearly 3% of projected water consumption for that year. If low-cost DAC can be deployed at scale in China, this could reduce bioenergy irrigation consumption to approximately 5 km$^3$ per year. But, evaporative losses from DAC itself are projected to consume large amounts of water, over 7 km³ per year in 2060, nearly 80 percent of municipal water consumption in China in 2015.

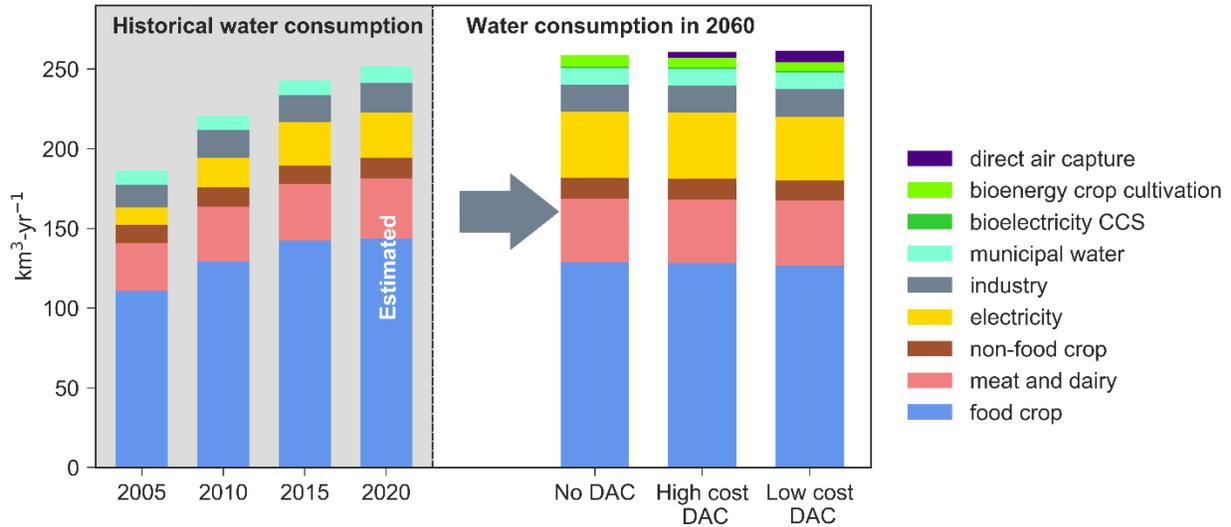

Figure 4: Recent historical and projected water consumption for China for net-zero $CO_2$ emissions by 2060.

Figure 5 reports historical and projected land-use for China if it and the rest of the world reach net-zero emissions by 2060, relative to a 2005 base year. The existing historical trend of gradual forest expansion is greatly accelerated, whereas grasslands see a reversal of recent historical growth as bioenergy and afforestation are scaled up. Land area for food cultivation (i.e., grains and staple crops) is projected to decline by nearly 100,000 km² from 2005 levels in China in 2060 in the no DAC scenario, while bioenergy crop production expands to over 190,000 km² and forested area gains approximately 430,000 km². Low-cost DAC can reduce the decline in food production land slightly, to approximately 90,000 km², with 148,000 km² being used for bioenergy crop cultivation, and 380,000 km² net gain in forest area from 2005 level. Without DAC available, weighted prices for major staple grain crops increase to 300% of their 2010 levels (200% increase) due to land competition from BECCS and afforestation. Low-cost DAC availability reduces this increase to approximately 175% of their 2010 levels (75% increase). Figure 5 also reports how the availability of DAC could reduce biomass cropland area required for mitigation and negative emissions in major river basins in China. In the most productive agricultural regions in eastern China, the availability of DAC could reduce biomass cropland area by 20-25%, freeing up more land for other agricultural activity or environmental conservation. Fractional land area devoted to biomass cropland in each water basin is reported in the Supplementary Information.

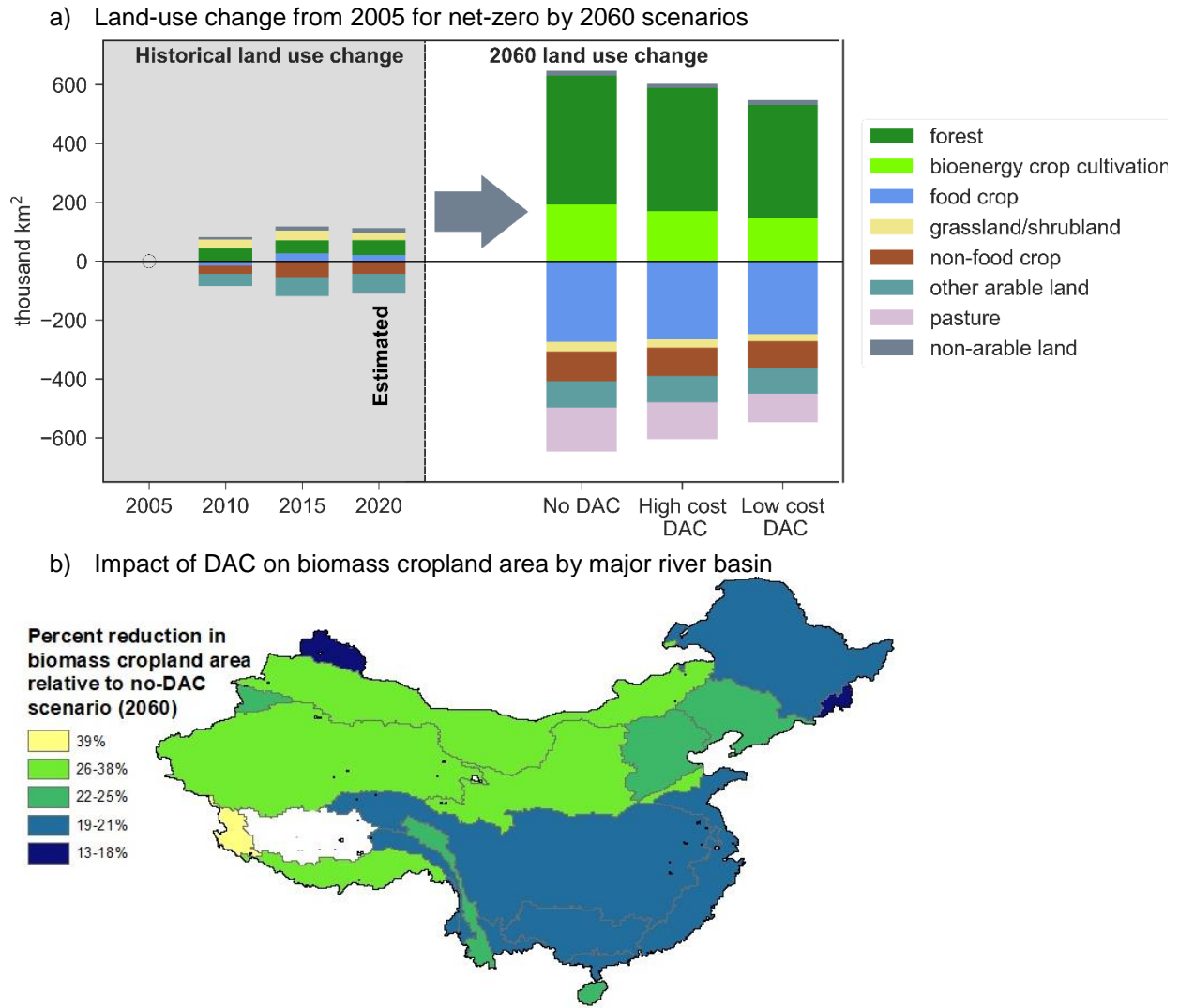

Figure 5: Historical and projected land-use changes from 2005 in China for net-zero $CO_2$ emissions by 2060 scenarios (a). Difference in biomass cropland area in major river basins in China between no DAC and low-cost DAC scenarios (b). White coloring indicates zero biomass cropland area in both scenarios and thus no change between them.

## 4. Sensitivity analysis

To more fully assess the sensitivity of our projections for DAC deployment in China to input assumptions, we individually varied the cost and energy input parameters for DAC itself, as well as other model assumptions that are external to DAC but directly influence its cost and/or requirement for its deployment. Table 3 summarizes the assumptions for our sensitivity analysis. Parametrizations for geologic carbon storage supply curves used in this sensitivity study, as well as population and GDP input assumptions for the "central" and "low residual emissions" scenarios may be found in the Supplementary Information. Figure 6 reports how parametric variations from Table 3 influence DAC deployment in China in 2060. All scenarios are compared against the "Low-cost DAC" scenario and hold all other input assumptions constant except for the indicated change. Assumptions for capital and non-energy operating costs have large impacts on the projected level of DAC deployment, with a 56% reduction in non-energy cost (i.e., from $180 to $78 in 2060) leading to over an 80% increase in deployment. Conversely, a 66% increase in non-energy cost (i.e., from $180 to $300) reduces the potential of DAC in China by approximately 40%. The role of DAC is reduced by over 60% if land available for bioenergy, afforestation, and agricultural expansion is unconstrained, which could be at odds with other objectives such as environmental conservation. The requirement for DAC could also be reduced by 45% if lower population growth and improved mitigation technologies reduce the amount of residual $CO_2$ that needs to be offset with DAC and other NETs. Increasing the geologic carbon storage cost by a factor of 10 reduces DAC deployment by 30%. This substantial but relatively small response to such a large parametric variation occurs due to the small contribution of storage to the overall cost of DAC, as well as because it similarly increases costs for abatement using fossil carbon capture and storage. Process heat and electrical input efficiencies remaining at the upper bounds of today's literature instead of improving over time could reduce the role of DAC by 20 and 10%, respectively.

Table 3. Assumptions for Sensitivity Analysis.

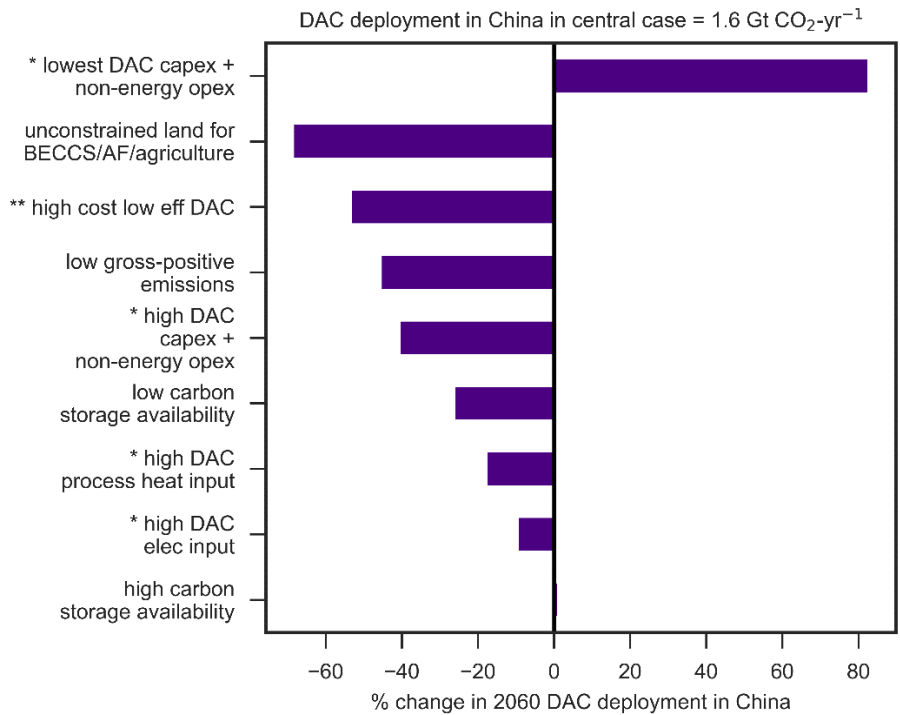

Figure 6: Sensitivity analysis for DAC deployment in China in 2060. Items marked with an asterisk (*) indicate direct changes to DAC input assumptions such as natural gas for process heat, electricity, and non-energy cost. The "high-cost low efficiency DAC" scenario is marked with two asterisks (**) and is the same as the "high-cost DAC" scenario in figures 2-5. Scenarios with no asterisks indicate changes to other GCAM assumptions that directly influence the cost and/or requirement for DAC.

## 4. Conclusions

In 2018, the IPCC released a report describing what it would take to limit end-of-century global warming to $+1.5^{o}C$ above preindustrial levels, recognizing this target would require ambition and coordination far beyond what we have seen to date.[53] The recent pledge from China to achieve carbon neutrality by 2060 provides a major contribution towards limiting climate change to below $+2^{o}C$ above pre-industrial levels. The rest of the world joining China's carbon neutrality pledge would help achieving the common goal of limiting climate warming to below $+2^{o}C$, but greater near-term emissions cuts and/or long-term future net-negative emissions would be required to meet the below $+1.5^{o}C$ goal. We modeled these different futures in an effort to understand the role of negative emissions technologies (NETs) in helping to achieve the net-zero by 2060 target. Negative emissions technologies are being developed to offset recalcitrant emissions from transportation and industry and potentially draw down atmospheric $CO_2$ levels in the future, even though they are generally considered to be more expensive than conventional decarbonization activities. We analyzed China's options to achieve net-zero $CO_2$ emissions using the Global Change Assessment Model (GCAM) with three major sources of negative emissions: bioenergy with carbon capture (BECCS), afforestation, and direct air capture with carbon storage (DAC).

Our findings show that DAC can play a large role in China to achieve its net-zero emissions target. This is particularly true if the rest of the world is decarbonizing alongside China. The extent to which DAC is deployed depends on its cost, which is expected to decrease over time as technologies improve with widespread implementation. However, some of the radiative forcing benefit of $CO_2$ removal with DAC is offset by fugitive methane emissions from its natural gas process heat requirement, and non-$CO_2$ emissions from increased residual fossil fuel use. Because of these effects, the global average temperature anomaly in 2100 is approximately $0.1^{\circ}C$ higher in the low-cost DAC scenario than in the no-DAC scenario, despite having the same $CO_2$ emissions trajectories. If the leakage rate of the natural gas supply chain is higher, the net radiative forcing benefit of large-scale natural gas-fired DAC deployment would be correspondingly lower. Without DAC, China could reach net-zero $CO_2$ emissions by 2060 with 1.5 $GtCO_2$ per year negative emissions, but this would need to come from BECCS and afforestation at an increasingly higher marginal cost per $tCO_2$ than if DAC could be deployed alongside them. With DAC widely available, China's carbon neutrality can be supported by more than 2-3 $GtCO_2$ per year negative emissions to get to the net-zero $CO_2$ emissions by 2060. Scaling up DAC to this level would require investment on the order of US\$ 200-280 billion per year in 2060, which is about 1-2% of China's GDP in 2019 and 0.5-0.7% of its projected GDP in 2060.

Our results indicate that up to 30-60% of China's negative emissions requirement could be fulfilled by DAC, with the remainder fulfilled by BECCS and afforestation. These results provide insight into the country-scale changes in water, energy, and land-use when different forms of negative emissions are used in China. We also assessed sensitivity of DAC deployment to key parameters such as its cost, energy efficiency, (lack of) constraints on land for BECCS and forest expansion, carbon storage costs, and ability to reduce gross-positive $CO_2$ emissions that need to be offset with DAC and other NETs. Improvements in DAC non-energy costs could increase its deployment in China by up to 80%. We also find that DAC still plays an important role even under more pessimistic assumptions regarding its future cost and energy efficiency improvements, as well as under more optimistic assumptions for future gross-positive emissions reductions. This is due to high costs of decarbonizing the industrial and freight transportation sectors. DAC availability could reduce biomass cropland area in major river basins in China by up to 25%, freeing up more land for agricultural production and environmental conservation. While DAC reduces the extent of land-use dedicated to climate mitigation (i.e., for afforestation and bioenergy), they remain large. DAC also reduces the need for irrigation water consumption for bioenergy crops, but itself consumes large amounts of water. This results in little change to overall water consumption in China if DAC is available, but DAC availability may be able to shift water use away from agricultural regions with growing water stress. The large remaining land and water tradeoffs of negative emissions with DAC availability result from its displacement of emissions abatement, which offsets the impacts of reducing the need for other forms of negative emissions. In addition to improvements which reduce the financial and natural resource intensity of DAC and other NETs themselves, greater investment in technologies to enable emissions avoidance such as transport and industrial electrification could also reduce the tradeoffs of negative emissions for offsets.

Acknowledgement:
The authors would like to acknowledge Katherine Holcomb of the UVA Advanced Research Computing Service for her assistance with setting up GCAM on UVA's High-Performance Computing Cluster.

Funding:
This work was supported by the University of Virginia's Office of the Vice President for Research—3 Cavaliers Program, the University of Virginia Environmental Resilience Institute, the Global Technology Strategy Program and the Alfred P. Sloan Foundation.

Data Availability: GCAM is an open-source integrated assessment model available at: https://github.com/JGCRI/gcam-core. The additional data underlying this article will be shared on reasonable request to the corresponding author.

# References

1.    Statement by H.E. Xi Jinping President of the People's Republic of China At the General Debate of the 75th Session of The United Nations General Assembly. https://www.fmprc.gov.cn/mfa_eng/zxxx_662805/t1817098.shtml.

2.    NDRC. Enhanced actions on climate change: China's intended nationally determined contributions. https://www4.unfccc.int/sites/ndcstaging/PublishedDocuments/China First/China%27s First NDC Submission.pdf (2015).

3.    Chen, W., Yin, X. & Zhang, H. Towards low carbon development in China: a comparison of national and global models. *Clim. Change* 136, 95–108 (2016).

4.    Zhou, N. *et al.* China's energy and emissions outlook to 2050: Perspectives from bottom-up energy end-use model. *Energy Policy* 53, 51–62 (2013).

5.    Chai, Q.-M. & Xu, H.-Q. Modeling an emissions peak in China around 2030:

Synergies or trade-offs between economy, energy and climate security. *Adv. Clim. Chang. Res.* 5, 169–180 (2014).

6.    Qi, Y. *Annual Review of Low-Carbon Development in China: 2010.* (World Scientific, 2013). doi:10.1142/8301.

7.    Zhou, S. *et al.* Energy use and CO2 emissions of China's industrial sector from a global perspective. *Energy Policy* 58, 284–294 (2013).

8.    Pan, X. *et al.* Implications of near-term mitigation on China's long-term energy transitions for aligning with the Paris goals. *Energy Econ.* 90, 104865 (2020).

9.    Fuhrman, J. *et al.* Food–energy–water implications of negative emissions technologies in a +1.5 °C future. *Nat. Clim. Chang.* 10, 920–927 (2020).

10.   Climate Action Tracker. Countries with submitted or proposed NDC updates. https://climateactiontracker.org/climate-target-update-tracker/ (2020).

11.   The Rt Hon Chris Skidmore MP. UK becomes first major economy to pass net zero emissions law - GOV.UK. *Department for Business, Energy & Industrial Strategy* https://www.gov.uk/government/news/uk-becomes-first-major-economy-to-pass-net-zero-emissions-law (2019).

12.   Statement attributable to the Spokesperson for the Secretary-General – on Japanese Prime Minister Suga's net-zero announcement | United Nations Secretary-General. https://www.un.org/sg/en/content/sg/statement/2020-10-26/statement-attributable-the-spokesperson-for-the-secretary-general-–-japanese-prime-minister-suga's-net-zero-announcement.

13.   UK net zero target | The Institute for Government. https://www.instituteforgovernment.org.uk/explainers/net-zero-target (2020).

14.   Peters, G. P., Davis, S. J. & Andrew, R. A synthesis of carbon in international trade. *Biogeosciences* 9, 3247–3276 (2012).

15.   Meng, J. *et al.* The rise of South-South trade and its effect on global CO2 emissions. *Nat. Commun.* 9, 1–7 (2018).

16.   Mi, Z., Meng, J., Green, F., Coffman, D. M. & Guan, D. China's "Exported Carbon" Peak: Patterns, Drivers, and Implications. *Geophys. Res. Lett.* 45, 4309–4318 (2018).

17.   Liu, Z. *et al.* Embodied carbon emissions in China-US trade. *Sci. China Earth Sci.* 63, 1577–1586 (2020).

18.   Zhang, Z. *et al.* Embodied carbon emissions in the supply chains of multinational enterprises. *Nat. Clim. Chang.* 10, (2020).

19.   CDIAC. Carbon Dioxide Information Analysis Center. https://cdiac.ess-dive.lbl.gov (2017).

20.   IEA. Global CO2 emissions in 2019 – Analysis. https://www.iea.org/articles/global-co2-emissions-in-2019 (2020).

21.   Wang, H. *et al.* Modeling of power sector decarbonization in China: comparisons of early and delayed mitigation towards 2-degree target. *Clim. Change* (2019) doi:10.1007/s10584-019-02485-8.

22.  Stanway, D. China's new coal power plant capacity in 2020 more than three times rest of world's: study | Reuters. https://www.reuters.com/article/us-china-coal/chinas-new-coal-power-plant-capacity-in-2020-more-than-three-times-rest-of-worlds-study-idUSKBN2A308U (2021).

23.  Myllyvirta, L., Zhang, S. & Shen, X. Analysis: Will China build hundreds of new coal plants in the 2020s? *Carbon Brief* https://www.carbonbrief.org/analysis-will-china-build-hundreds-of-new-coal-plants-in-the-2020s (2020).

24.  Monitor, G. E. & Air, C. for R. on E. and C. *China Dominates 2020 Coal Plant Development.* (2021).

25.  Wang, H. & Chen, W. Modelling deep decarbonization of industrial energy consumption under 2-degree target: Comparing China, India and Western Europe. *Appl. Energy* 238, 1563–1572 (2019).

26.  Yin, X. *et al.* China's transportation energy consumption and CO2 emissions from a global perspective. *Energy Policy* 82, 233–248 (2015).

27.  Zhang, H., Chen, W. & Huang, W. TIMES modelling of transport sector in China and USA: Comparisons from a decarbonization perspective. *Appl. Energy* 162, 1505–1514 (2016).

28.  Pietzcker, R. C. *et al.* Long-term transport energy demand and climate policy: Alternative visions on transport decarbonization in energy-economy models. *Energy* 64, 95–108 (2014).

29.  Le Quéré, C. *et al.* Temporary reduction in daily global CO2 emissions during the COVID-19 forced confinement. *Nat. Clim. Chang.* 10, 647–653 (2020).

30.  Le Quéré, C. *et al.* Fossil CO2 emissions in the post-COVID-19 era. *Nat. Clim. Chang.* 11, 197–199 (2021).

31.  Yu, S. *et al. Five strategies to achieve china's 2060 carbon neutrality goal.* https://www.efchina.org/Reports-en/report-lceg-20200929-en (2020).

32.  Zheng, X. *et al.* Drivers of change in China's energy-related CO2 emissions. *Proc. Natl. Acad. Sci. U. S. A.* 117, 29–36 (2020).

33.  Fuss, S. *et al.* Negative emissions—Part 2: Costs, potentials and side effects. *Environ. Res. Lett.* 13, 063002 (2018).

34.  Cao, S. Why large-scale afforestation efforts have failed to solve the desertification problem. *Environ. Sci. Technol.* 1826–1831 (2008).

35.  Cao, S. *et al.* Excessive reliance on afforestation in China's arid and semi-arid regions: Lessons in ecological restoration. *Earth-Science Reviews* vol. 104 240–245 (2011).

36.  Jia, X., Shao, M., Zhu, Y. & Luo, Y. Soil moisture decline due to afforestation across the Loess Plateau, China. *J. Hydrol.* 546, 113–122 (2017).

37.  Wang, J. *et al.* Large Chinese land carbon sink estimated from atmospheric carbon dioxide data. *Nature* 586, 720–723 (2020).

38.  Fuhrman, J., McJeon, H., Doney, S. C., Shobe, W. & Clarens, A. F. From Zero to Hero?: Why Integrated Assessment Modeling of Negative Emissions Technologies

Is Hard and How We Can Do Better. *Front. Clim.* 1, 11 (2019).

39. NRC. Negative Emissions Technologies and Reliable Sequestration. *Natl. Acad. Press* (2018) doi:10.17226/25259.

40. Zastrow, M. China's tree-planting drive could falter in a warming world. *Nature* vol. 573 474–475 (2019).

41. Muratori, M., Calvin, K., Wise, M., Kyle, P. & Edmonds, J. Global economic consequences of deploying bioenergy with carbon capture and storage (BECCS). *Environ. Res. Lett.* 11, (2016).

42. Popp, A. *et al.* Land-use futures in the shared socio-economic pathways. *Glob. Environ. Chang.* 42, 331–345 (2017).

43. Wise, M. *et al.* Implications of Limiting CO2 Concentrations for Land Use and Energy. *Science (80-. ).* 324, 1183–1186 (2009).

44. Smith, P. *et al.* Biophysical and economic limits to negative CO2emissions. *Nat. Clim. Chang.* 6, 42–50 (2016).

45. US Department of Commerce, N. G. M. L. Global Monitoring Laboratory - Carbon Cycle Greenhouse Gases.

46. Microsoft. Progress on our goal to be carbon negative by 2030. https://blogs.microsoft.com/on-the-issues/2020/07/21/carbon-negative-transform-to-net-zero/ (2020).

47. ExxonMobil and Global Thermostat to Advance Breakthrough Atmospheric Carbon Capture Technology | Business Wire. https://www.businesswire.com/news/home/20190627005137/en/ExxonMobil-Global-Thermostat-Advance-Breakthrough-Atmospheric-Carbon.

48. Chevron, Occidental invest in CO2 removal technology - Reuters. https://www.reuters.com/article/us-carbonengineering-investment/chevron-occidental-invest-in-co2-removal-technology-idUSKCN1P312R.

49. United Airlines invests in carbon-capture project to be 100% green by 2050 | Reuters. https://www.reuters.com/article/united-arlns-climate-occidental/united-airlines-invests-in-carbon-capture-project-to-be-100-green-by-2050-idUSKBN28K1NE.

50. Carbon Engineering inks Shopify as its first partner for carbon removal as a service | TechCrunch. https://techcrunch.com/2021/03/09/carbon-engineering-inks-shopify-as-its-first-partner-for-carbon-removal-as-a-service/.

51. Yu, S. *et al.* CCUS in China's mitigation strategy: insights from integrated assessment modeling. *Int. J. Greenh. Gas Control* 84, 204–218 (2019).

52. Rogelj, J. *et al.* Mitigation pathways compatible with 1.5°C in the context of sustainable development. *Spec. Report, Intergov. Panel Clim. Chang.* (2018).

53. Rogelj, J. *et al.* Mitigation Pathways Compatible with 1.5$\degree$C in the Context of Sustainable Development. *Spec. Report, Intergov. Panel Clim. Chang.* (2018).

54. IPCC. *IPCC Special Report on 1.5 C, Chapter 2.* (2018).

55. Realmonte, G. *et al.* An inter-model assessment of the role of direct air capture in deep mitigation pathways. *Nat. Commun.* 10, 3277 (2019).

56. Fuhrman, J. *et al.* Food–Energy–Water Implications of Negative Emissions Technologies in a +1.5\$degree\$C Future. *Nat. Clim. Chang.* 10, 920–927 (2020).

57. Strefler, J. *et al.* Between Scylla and Charybdis: Delayed mitigation narrows the passage between large-scale CDR and high costs. *Environ. Res. Lett.* 13, 44015 (2018).

58. Chen, C. & Tavoni, M. Direct air capture of $CO_2$ and climate stabilization: A model based assessment. *Clim. Change* 118, 59–72 (2013).

59. JGCRI. GCAM v5.3 documentation: Global Change Analysis Model (GCAM). https://jgcri.github.io/gcam-doc/ (2020).

60. Kaufman, N., Barron, A. R., Krawczyk, W., Marsters, P. & McJeon, H. A near-term to net zero alternative to the social cost of carbon for setting carbon prices. *Nat. Clim. Chang.* (2020) doi:10.1038/s41558-020-0880-3.

61. Bednar, J., Obersteiner, M. & Wagner, F. On the financial viability of negative emissions. *Nat. Commun.* 10, 1783 (2019).

62. Daniel, K. D., Litterman, R. B. & Wagner, G. Declining CO 2 price paths . *Proc. Natl. Acad. Sci.* 201817444 (2019) doi:10.1073/pnas.1817444116.

63. Smith, P. *et al.* The marker quantification of the Shared Socioeconomic Pathway 2: A middle-of-the-road scenario for the 21st century. *Glob. Environ. Chang.* 42, 251–267 (2017).

64. Wise, M. *et al.* Implications of limiting $CO_2$ concentrations for land use and energy. *Science* 324, 1183–6 (2009).

65. Keith, D. W., Holmes, G., St. Angelo, D. & Heidel, K. A process for capturing $CO_2$ from the atmosphere. *Joule* 2, 1573–1594 (2018).

66. Zeman, F. Energy and material balance of $CO_2$ capture from ambient air. *Environ. Sci. Technol.* 41, 7558–7563 (2007).

67. Stolaroff, J. K., Keith, D. W. & Lowry, G. V. Carbon dioxide capture from atmospheric air using sodium hydroxide spray. *Environ. Sci. Technol.* 42, 2728–2735 (2008).

68. Fasihi, M., Efimova, O. & Breyer, C. Techno-economic assessment of $CO_2$ direct air capture plants. *J. Clean. Prod.* 224, 957–980 (2019).

69. Realmonte, G. *et al.* An Inter-model Assessment of the Role of Direct Air Capture in Deep Mitigation Pathways. *Nat. Commun.* 10, 3277 (2019).

70. Le Quéré, C. *et al.* Global Carbon Budget 2018. *Earth Syst. Sci. Data* 10, 2141–2194 (2018).

71. US Department of Commerce, NOAA, E. S. R. L. ESRL Global Monitoring Division - Global Greenhouse Gas Reference Network. *NOAA Res.* (2019).

72. Environment European Agency. Global average near surface temperatures relative to the pre-industrial period. https://www.eea.europa.eu/data-and-maps/daviz/global-average-air-temperature-anomalies-5#tab-dashboard-02 (2019).

Table 1. Description of the $CO_2$ emissions constraint trajectories run in GCAM to understand the role DAC in meeting these constraints

| Scenario Name | Description |
| --- | --- |
| No climate policy | A reference scenario with no climate mitigation policy (i.e., pricing or constraints on $CO_2$ or other GHG emissions), but improving technological efficiency |
| China + rest of the world (ROW) net-zero 2060 | China, along with the rest of the world, achieve net-zero $CO_2$ by 2060 in a linearly declining emissions trajectory. China's emissions are individually constrained, but emissions from the remaining regions in the rest of the world are allowed to be individually greater than or less than a separately imposed emissions constraint, so long as their sum is less than or equal to this constraint. |

Table 2. Input parameters for DAC technology[56,65]

| Technology | Natural Gas (GJ/tCO$_2$) | | Electricity (GJ/tCO$_2$) | | Non-Energy Cost (2015 \$/tCO$_2$) | | Water (m$^3$/tCO$_2$) |
|---|---|---|---|---|---|---|---|
| | 2020 | 2050 | 2020 | 2050 | 2020 | 2050 | - |
| Low cost DAC | 8.1 | 5.3 | 1.8 | 1.3 | 300 | 180 | 4.7 |
| High cost DAC | 8.1 | 8.1 | 1.8 | 1.8 | 300 | 300 | 4.7 |

Note: For low-cost DAC, we assume that the energy efficiency of the technology improves between 2020 and 2050. The energy and non-energy cost inputs are assumed decline linearly between year 2020 and 2050 and thereafter remain constant. For the high cost DAC scenario, we assume that the technology will remain costly and energy-intensive.

Table 3. Assumptions for Sensitivity Analysis.

| Scenario Name | Description |
|---|---|
| Central | DAC energy and cost inputs follow the trajectory defined in the "Low cost DAC" scenario (Table 2). Socioeconomic assumptions follow the "middle of the road" scenario in the GCAM core release. 90% of non-commercial land is assumed protected from agricultural expansion. |
| Best-case DAC capex + non-energy opex | Non-energy cost inputs decline linearly to $78 per $tCO_2$, the most optimistic value found in today's literature[65] |
| High DAC capex + non-energy opex | Non-energy cost inputs do not improve between 2020 and 2060, remaining at $300 per $tCO_2$ |
| High DAC process heat input | Natural gas process heat requirement for DAC does not improve between 2020 and 2060, remaining at 8.1 GJ per $tCO_2$ |
| High DAC electricity input | Electricity input requirement for DAC does not improve between 2020 and 2060, remaining at 1.8 GJ per $tCO_2$ |
| Unconstrained land for BECCS/afforestation/agriculture | Removed 90% protection constraint on non-commercial lands, freeing up more land for bioenergy, forestry, and other agricultural activity. |
| Low carbon storage availability | Cost curves for geologic carbon storage follow the "highest cost CCS" assumption from the GCAM core release. Offshore carbon storage reservoirs are assumed unavailable. |
| High carbon storage availability | Cost curves for geologic carbon storage follow the highest availability of CCS assumed in the GCAM core release. |
| Low residual emissions | Population growth, technological improvement, and social preferences generally follow the "sustainable development" scenario. Geologic carbon storage supply curves and land protection constraints are the same as in the central scenario. |

Note: all other input assumptions other than the one described in each scenario are held constant, including the $CO_2$ emissions constraint.